\documentclass[%
 aip,
rsi,%
 amsmath,amssymb,
preprint,%
]{revtex4-1}

\usepackage{graphicx}
\usepackage{dcolumn}
\usepackage{bm}
\usepackage[utf8]{inputenc}
\usepackage[T1]{fontenc}
\usepackage{mathptmx, mathrsfs}
\usepackage{enumitem}
\usepackage{verbatim}
\usepackage{xcolor}
\newlist{steps}{enumerate}{1}
\setlist[steps, 1]{label = Step \arabic*:}
\usepackage{listings}
\newcommand{\bra}[1]{\ensuremath{\left\langle#1\right|}}
\newcommand{\ket}[1]{\ensuremath{\left|#1\right\rangle}}

\newcommand{\bracket}[2]{\ensuremath{\left\langle #1 \middle| #2 \right\rangle}}

\begin{document}

\preprint{}

\title[]{A Robust and Unified Solution for Choosing the Phases of Adiabatic States as a Function of Geometry: Extending Parallel Transport Concepts to the cases of Trivial \& Near Trivial Crossings}

\author{Zeyu Zhou}
 
 \affiliation{Department of Chemistry, University of Pennsylvania, Philadelphia, Pennsylvania 19104, United States}
 
\author{Zuxin Jin}
\author{Tian Qiu}

\author{Andrew M. Rappe}

\author{Joseph Eli Subotnik}
 \email{subotnik@sas.upenn.edu}
 \affiliation{Department of Chemistry, University of Pennsylvania, Philadelphia, Pennsylvania 19104, United States}

\date{\today}

\begin{abstract}
We investigate a simple and robust scheme for choosing the phases of adiabatic electronic states smoothly (as a function of geometry) so as to maximize the performance of ab initio non-adiabatic dynamics methods. Our approach is based upon consideration of the overlap matrix ($\mathbf{U}$) between basis functions at successive points in time and selecting the phases so as to minimize the matrix norm of $\log(\mathbf{U})$. In so doing, one can extend the concept of parallel transport to cases with sharp curve crossings. 
We demonstrate that this algorithm performs well under extreme situations where dozens of states cross each other either through trivial crossings (where there is zero effective diabatic coupling), or through nontrivial crossings (when there is a nonzero diabatic coupling), or through a combination of both.
In all cases, we compute the time-derivative coupling matrix elements (or equivalently non-adiabatic derivative coupling matrix elements) that are as smooth as possible.
Our results should be of interest to all who are interested in either non-adiabatic dynamics, or more generally, parallel transport in large systems.
\end{abstract}
\maketitle
\section{\label{sec:intro}Introduction}
\subsection{Parallel transport in the non-adiabatic regime}
For many problems in chemical physics, one must naturally deal with a quantum subsystem that evolves either in time or according to some external parameter. For instance, the Born-Oppenheimer approximation requires that one consider quantum electronic states as parametrized by nuclear geometry; the eigenvalues of the electronic Schrodinger equation become the potential energy surfaces which are the bedrock of modern chemistry. More generally, however, there is also a long story going back to Longuet-Higgins\cite{longuet1958studies}, Baer\cite{baer1975adiabatic} and Mead and Truhlar\cite{mead1979determination} pointing out that when we make the Born-Opphenheimer approximation, we should not focus exclusively on how quantum \textit{eigenvalues} depend on external parameters, but also on how the \textit{eigenvectors} themselves evolve. In a famous set of papers considering how electronic states depend on the vector potential or magnetic field, Berry showed that the phases of the eigenvectors can have a great deal of rich, topological physics buried within them.\cite{berry1984quantal} 
More specifically, suppose one propagates a single eigenvector slowly around a closed loop parametrized by $t\in[0, T]$. Berry showed that in the adiabatic limit – which means that two eigenvalues are never close to each other – that eigenvector picks up an extra phase (beyond the dynamical phase) as a function of the classical parameter it depends on.\cite{simon1983holonomy, berry1984quantal, yarkony1996diabolical, yarkony1996current, baer2006beyond} Thus, Berry's phase demonstrated the limits of so-called \textit{parallel transport} -- which means that $\bracket{\phi_{j}(t)}{\phi_{j}(t+dt)}$ is real and maximized, so that one would expect $\bracket{\phi_{j}(t)}{\dot{\phi}_{j}(t+dt)}\approx 0$; even if one parametrizes adiabatic states according to parallel transport, an adiabatic wavefunction picks up a phase when the external parameter traces a complete cycle. Therefore, from a different point of view, Berry demonstrated that parallel transport is not consistent with the presence of globally well-defined adiabatic state (with globally well-defined phases).

Now, despite the failure of parallel transport to account for a global, topological Berry phase, for many practical purposes, parallel transport works well enough and can solve many interesting problems. For instance, in the context of non-adiabatic molecular dynamics, parallel transport is always applied as one propagates nuclear trajectories that explore different nuclear geometries in time; one wants a smooth choice of adiabatic states, and one usually does not ever return to the initial location. In such a case, if one never completes a closed cycle, ignoring Berry's phase (i.e. $\bracket{\phi_{j}(t)}{\dot{\phi}_{j}(t+dt)}\approx 0$) is usually a well-behaved, efficient and accurate approximation.\footnote{In fact, noting that, in practice, calculating $T_{jj} = \bracket{\phi_{j}}{\dot{\phi}_{j}}$ is very difficult or impossible for almost all ab initio calculations, one can argue that parallel transport is not only sensible but actually required for many modern programs.} Nevertheless, for many problems in chemical physics, implementing parallel transport is not obvious in practice. After all, let $\mathbf{U}$ be the overlap matrix between electronic states:
\begin{eqnarray}
U_{jk} = \bracket{\phi_{j}(t)}{\phi_{k}(t+dt)}.
\label{eq: overlap_matrix}
\end{eqnarray}
Then, the usual parallel transport approach tells us to make all diagonal elements $U_{jj}$ real and positive, which fixes the sign of each $\ket{\phi_{j}(t)}$ at each time step. This phase convention is used nowadays routinely for modern ab initio non-adiabatic dynamics calculations (Ehrenfest, FSSH,\cite{tully1990molecular, zaari2015nonadiabatic} or AIMS\cite{ben2000ab}) to investigate photo-excited relaxation. And yet, if one moves away from the adiabatic regime, insisting that $U_{jj}$ are real and maximized can be difficult or even unstable: what if $U_{jj}\approx 0$ as might be possible in the non-adiabatic regime?

To better understand this failure conceptually, consider how one would apply the concept of parallel transport in the extreme non-adiabatic limit of curve crossings, also called a trivial crossing.\cite{hammes1994proton, fabiano2008implementation, barbatti2010non, meek2014evaluation, wang2014simple, wang2016recent,  jain2016efficient, lee2019solving} Consider a two-level model Hamiltonian:
\[\mathbf{H}_{\rm{real}} = \left[
\begin{array}{c c }
0.1 \tanh(R) & \kappa\exp(-R^2)\\
\kappa\exp(-R^2) & -0.1 \tanh(R)  \\
\end{array} \right]
\]
\begin{figure}
    \centering
    \includegraphics[width=0.5\textwidth]{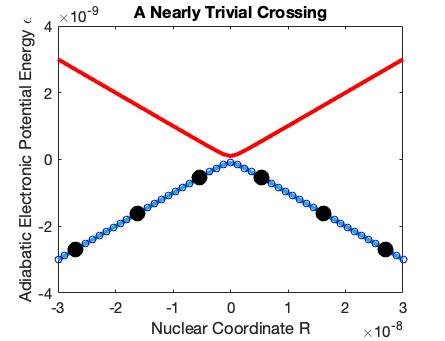}
    \caption{A schematic diagram for a nearly trivial crossing. The red and blue lines represent two adiabatic states with a small gap between them at R=0.  The small circles demonstrate that, if we are to rely on parallel transport to determine the phase of the wavefunctions at each successive point in space, we will need a very dense set of  grid points to model the curve crossing at R=0. By contrast, we would like to run dynamics with a larger time step and a more spare sampling of grid points, e.g. the set of black solid circles. In such a case, we will need a clever algorithm (beyond simple parallel transport) to pick the phases of the adiabatic states because the set of such (black circle) points will miss crucial details of the near trivial curve crossing.}
    \label{fig: TrivialCrossing}
\end{figure}
There is an avoided crossing at $R = 0$, if $\kappa \neq 0$, and let us choose $\kappa$ to be very small, e.g. $1\times10^{-10}$. As shown in Fig.~{\ref{fig: TrivialCrossing}}, if we strictly follow parallel transport, we will need many such grid points (as labeled by small circles) to transport our states, and thus a very small time step will be required for the simulation; the computational cost for simulating a curve crossing will be unbearably and unnecessarily large (and without much physical meaning). In practice, we would actually prefer a large time step (as shown by the black circles), and simply recognize that two states switched. But how to choose phases? The overlap matrix $\mathbf{U}$ between the two central black circles in Fig. ~\ref{fig: TrivialCrossing} will take the following form
\[\mathbf{U}_{\rm{real}} = \left[
\begin{array}{c c }
0 & \pm1 \\
\pm1 & 0  \\
\end{array} \right]
\]
The $+/-$ signs should in principle be determined by small time steps and true parallel transport, but if we need not care, can we avoid all the cost?

While the situation above may appear artificial, the basic premise of setting phases so as to make the diagonal matrix elements real and maximally positive can be also problematic in less obvious cases. Consider the following overlap matrix $\mathbf{U}_{pt}$ that obeys parallel transport in $N$-dimensions:
\begin{eqnarray}
\mathbf{U}_{pt} = \left[
\begin{array}{c c c c c }
1 - \frac{2}{N} & -\frac{2}{N} & -\frac{2}{N} & \dots & -\frac{2}{N}  \\
-\frac{2}{N} & 1 - \frac{2}{N} & -\frac{2}{N} & \dots & -\frac{2}{N} \\
-\frac{2}{N} & -\frac{2}{N} & 1 - \frac{2}{N} & \dots & -\frac{2}{N} \\
\vdots & \vdots & \vdots & \ddots & \vdots \\
-\frac{2}{N} & -\frac{2}{N} & -\frac{2}{N} & \dots & 1 - \frac{2}{N} \\
\end{array} \right]
\label{eq: mat_Upt}
\end{eqnarray}
This matrix is unitary and as $N\rightarrow\infty$, the off-diagonal matrix elements approach zero, and the diagonal matrix elements approach unity. Nevertheless, $\det(\mathbf{U_{pt}}) = -1$, \footnote{A simple proof is as follows: set $\vec{a}$ to be a $1 * N$ vector with all elements +1. Clearly, $\mathbf{U}_{pt} = \mathbf{I}_{N} - \frac{2}{N}\vec{a}^{T}\vec{a}$. By using the Weinstein–Aronszajn identity, we immediately obtain $\det(\mathbf{U}_{pt}) = \det(\mathbf{I}_{N} - \frac{2}{N}a^{T}a) = \det(I_{1} -\frac{2}{N} aa^{T}) = 1-\frac{2}{N}N = -1$.} and so, even choosing the diagonal matrix elements to be near $+1$ is not necessarily a good extension of parallel transport (see Section~\ref{subsec: det1}).

Thus, we are clearly in need of a \textit{new} protocol to extend parallel transport to the non-adiabatic regime for arbitrary (and not necessarily diagonally dominant) overlap matrices $\mathbf{U}$. At this point, a mathematically-inclined reader might ask: is it at all reasonable or even possible to make such an extension without taking very small steps? Unfortunately, from the chemists’ perspective, there is no known alternative. Many dynamical approaches (especially FSSH and AIMS) are clearly optimal in an adiabatic basis and simple numerical tests can clearly show that choosing phases incorrectly can lead to very bad results. And this failure is not surprising: the choice of phase carries dynamical information quantum mechanically. In general, choosing $U_{jj}$ to be real and maximized is a good idea, since this process forces each eigenvector to change as slowly as possible without distorting the motion of the nuclear degrees of freedom, such that the adiabatic representation becomes a strong framework for smoothly expanding the nuclear Schr\"{o}dinger equation. In particular, one's choice of $\mathbf{U}$ defines one's choice of the non-adiabatic derivative coupling $\vec{\mathbf{d}}$, which is the key matrix that breaks the Born-Oppenheimer approximation. In turn, $\vec{\mathbf{d}}$ is critically dependent on the choices of phases of $\mathbf{U}$. Thus, choosing $\mathbf{U}$ wisely can enforce smoothness of $\vec{\mathbf{d}}$.

With this sensitivity in mind, there is no choice: for chemical dynamics, we must find a stable and optimal answer to the question -- what criteria should be selected for choosing the phases of $\mathbf{U}$? Alternatively, if significant improvement is not possible, we will be forced to take very small time steps.
\subsection{Choosing adiabatic state phases specifically in the context of non-adiabatic dynamics}
At this point, let us consider the case of non-adiabatic dynamics more explicitly. In this context, $\mathbf{U}$ is of paramount importance because it is related to the time-averaged derivative coupling $\vec{\mathbf{d}}$, which is used to propagate the equations of motion in the electronic degrees of freedom.
\begin{eqnarray}
i\hbar \dot{c}_{j} = H^{el}_{jj}c_{j} - i\hbar\sum_{k}\vec{d}_{jk}\cdot \vec{v}c_{k}
\label{eq: TDESE}
\end{eqnarray}
Here, $\vec{d}_{jk}$ is defined by
\begin{eqnarray}
\vec{d}_{jk} = \bracket{\phi_{j}^{ad}(\vec{R}(t))}{\vec{\nabla}_{\vec{R}}\phi_{k}^{ad}(\vec{R}(t + dt))}
\label{eq: ddddcccc}
\end{eqnarray}

Of course, chemists have long known that propagating Eqs.~(\ref{eq: TDESE}) and (\ref{eq: ddddcccc}) together is a bad idea. After all, $\vec{d}_{jk}$ is usually obtained by Hellmann-Feynman theorem as:
\begin{eqnarray}
\vec{d}_{jk} = \frac{\bra{\phi_{j}^{ad}}\vec{\nabla}_{\vec{R}}H_{el}(\vec{R})\ket{\phi_{k}^{ad}}}{\epsilon_{k} - \epsilon_{j}}
\end{eqnarray}
Thus, $\vec{d}_{jk}$ explodes when there is a crossing between adiabatic states $j$ and $k$ ($\vec{d}_{jk} \rightarrow \infty$ as $(\epsilon_{k} - \epsilon_{j})\rightarrow 0$), and one will need very small time steps near a crossing. With this limitation in mind, one better scheme is to take the logarithm of $\mathbf{U}$, by which the time-averaged time-derivative coupling matrix $\mathbf{T}$ is obtained.\cite{jain2016efficient}
\begin{eqnarray}
\frac{1}{dt}\int_{t}^{t+dt}d\tau\vec{d}_{jk}(\tau) \cdot \vec{v}(\tau) = T_{jk}(t + dt/2) = \bracket{\phi_{j}^{ad}}{\frac{d\phi_{k}^{ad}}{dt}} = \frac{[\log(\mathbf{U})]_{jk}}{dt}.
\label{eq:TlogU}
\end{eqnarray} 
As argued by Meek and Levine in a two-state context, \cite{meek2014evaluation} this approach is equivalent to calculating the time-averaged derivative coupling and can be used easily in dynamics (and is more stable than the Hammes-Schiffer-Tully method.\cite{hammes1994proton})

Nevertheless, to implement the Meek-Levine approach\cite{meek2014evaluation, jain2016efficient} or other existing trivial crossing approaches,\cite{hammes1994proton, fabiano2008implementation, wang2014simple, wang2016recent, lee2019solving} the inevitable question remains: what are the {\em phases} of $\mathbf{U}$?
\cite{footnote112, granucci2012surface, plasser2012surface} And, for the skeptical reader who thinks that choosing +/- phases cannot be very important, consider this: what if the Hamiltonian is complex? There will be an entire manifold of possible choices of phases. Specifically at trivial crossing, we expect an overlap matrix $U_{\rm{complex}}$ to be of the form:
\[\mathbf{U}_{\rm{complex}} = \left[
\begin{array}{c c }
0 & e^{i\theta} \\
-e^{-i\theta} & 0  \\
\end{array} \right]\]
and there will be infinitely many possible choices for $\theta$, each of which implies different physics for coupled nuclear-electron motion, and choosing an incorrect or inconsistent $\theta$ can lead to catastrophic consequences even for real $\mathbf{U}$ (where $\theta=0$ or $\pi$).\cite{akimov2018simple}

Obviously, for future purposes, we require a robust solution for picking the phases of the columns of $\mathbf{U}$, one that is applicable in the adiabatic and non-adiabatic limits. 
\subsection{Outline}
In this paper, we will extend the concept of parallel transport to the non-adiabatic regime by choosing the phases of $\mathbf{U}$ such that for each time step, $\mathbf{U}$ is a proper rotation matrix, no matter whether there is one trivial crossing, multiple trivial crossings or no trivial crossings, in both the real and complex regimes. In Section~{\ref{sec:2}}, we will introduce algorithms for both the real regime (which is easy) and for the complex regime (which is a bit harder). In Section~{\ref{sec:3}}, we will test our prescription on a modified model problem. In Section~{\ref{sec:4}}, we conclude and make several observations about the algorithm.

\section{Method \label{sec:2}}
Consider a path in configuration space, and let $\mathbf{U}$ be the overlap matrix between adiabatic basis sets at different times, as defined in Eq.~({\ref{eq: overlap_matrix}}). Note that $\mathbf{U}$ converges to the identity when the time step $dt$ approaches 0. Of course, for an identity matrix, the logarithm is a null matrix, and so one can argue that the optimal phases of $\mathbf{U}$ should be those phases that minimize the norm of all elements of $\mathbf{T}$. Unfortunately however, even for the case of real $\mathbf{U}$, it would be very expensive to directly calculate the $\mathbf{T}$ matrices for all possible signs of the columns of $\mathbf{U}$: in principle one would require $2^N$ logarithm calculations (i.e. matrix diagonalizations) so as to find the matrix $\mathbf{U}$ with the optimal sign choices. Moreover, in the complex regime, it would be impossible to minimize $\sum_{jk}|T_{jk}|^2$ and find the truly optimal $\mathbf{U}$ without having a grid in $\theta$ as well (which would require {$(N_{\theta})^N$ diagonalizations). One would prefer a reasonable set of approximations for minimizing $\sum_{jk}|T_{jk}|^2$ (or equivalently $\text{Tr}(|\log(\mathbf{U})|^2)$).
\subsection{Necessary Conditions for Minimizing $\text{Tr}(|\log(\mathbf{U})|^2)$}
\subsubsection{We must insist $\det(\mathbf{U}) = 1$ \label{subsec: det1}}
Let us now show that one of the necessary conditions for minimizing $\text{Tr}(|\log(\mathbf{U})|^2)$ is that $\det(\mathbf{U})$ should be $+1$.

For any unitary $\mathbf{U}$, we can decompose $\mathbf{U}$ as:
\begin{eqnarray}
\mathbf{U} = \mathbf{R}\boldsymbol{\Lambda} \mathbf{R}^{\dagger}
\label{eq: decomposition}
\end{eqnarray}
where $\boldsymbol{\Lambda}$ is diagonal and takes the following form:
\begin{equation}
\boldsymbol{\Lambda} = \left[
\begin{array}{c c c c c }
e^{i\delta_{1}} &  &  &  &   \\
 & e^{i\delta_{2}} &  &  & \\
 & & \ddots & & \\
 & & & e^{i\delta_{N-1}} & \\
 & & & & e^{i\delta_{N}} \\
\end{array} \right]
\label{eq: mat_diag}
\end{equation}
Suppose we change $\mathbf{U}$ to $\mathbf{U}'$ by
\begin{eqnarray}
\mathbf{U}' = e^{i\alpha}\mathbf{U}
\end{eqnarray}
the resulting $\mathbf{U}'$ is still unitary and can be expanded as,
\begin{eqnarray}
\mathbf{U}' = \mathbf{R}\Lambda' \mathbf{R}^{\dagger}
\end{eqnarray}
where
\[
\boldsymbol{\Lambda}' = e^{i\alpha}\boldsymbol{\Lambda} = \left[
\begin{array}{c c c c c }
e^{i(\delta_{1}+ \alpha)} &  &  &  &   \\
 & e^{i(\delta_{2} + \alpha)} &  &  & \\
 & & \ddots & & \\
 & & & e^{i(\delta_{N-1} + \alpha)} & \\
 & & & & e^{i(\delta_{N} + \alpha)} \\
\end{array} \right]
\]
For $\mathbf{U}'$, the quantity $\text{Tr}(|\log(\mathbf{U'})|^2)$ is
\begin{eqnarray}
\text{Tr}(|\log(\mathbf{U'})|^2) = \sum_{N}(\delta_{N} + \alpha)^2
\end{eqnarray}
One necessary condition for the quantity to be at minimum is 
\begin{eqnarray}
\frac{\partial\text{Tr}(|\log(\mathbf{U'})|^2)}{\partial\alpha}= \sum_{N}2(\delta_{N} + \alpha) = 0
\end{eqnarray}
This implies that, if U is the exact minimum, we must have
\begin{eqnarray}
\sum_{N}\delta_{N} = 0
\end{eqnarray}
Immediately,
\begin{eqnarray}
\det(\mathbf{U}) = e^{i\sum_{N}\delta_{N}} = 1
\end{eqnarray}
Hence, one necessary condition for the quantity $\text{Tr}(|\log(\mathbf{U})|^2)$ to be minimized is that $\det(\mathbf{U}) \equiv 1$. However, this is far from a sufficient condition.
\subsubsection{We must minimize a polynomial function of $\mathbf{U}$\label{subsubsec: min_poly_U}}
We can find another approximate, necessary condition besides the constraint $\det(\mathbf{U}) = 1$ by expanding $\log(\mathbf{U})$ into a Taylor series and truncating at second order around the identity.
Note that this approximation represents a slightly dangerous approach, because we are interested in the phases of adiabatic states around trivial or near trivial crossings, where $\mathbf{U}$ is far from the identity, and there is no reason to presume that a Taylor series around the identity should converge; and a second-order approximation need not be accurate at all. Nevertheless, bearing in mind this caveat, we will proceed and test this approximation numerically. To second order, 
\begin{eqnarray}
\text{Tr}(|\log(\mathbf{U'})|^2) &\approx& \sum_{jk} ((\mathbf{U}-\mathbf{I}) - (\mathbf{U}-\mathbf{I})^2 / 2)_{jk}((\mathbf{U}^*-\mathbf{I}) - (\mathbf{U}^*-\mathbf{I})^2 / 2)_{jk} \nonumber\\
&=&\sum_{jk} \Big(- \frac{\mathbf{U}^2}{2} + 2\mathbf{U} - \frac{3}{2}\mathbf{I}\Big)_{jk}\Big(- \frac{(\mathbf{U}^{*})^2}{2} + 2\mathbf{U}^{*} - \frac{3}{2}\mathbf{I}\Big)_{jk}
\label{eq:traceofTsquared}
\end{eqnarray}
Since $\mathbf{U}$ is a unitary matrix, i.e.
\begin{eqnarray}
\mathbf{U}\mathbf{U}^{\dagger} = \mathbf{U}^{\dagger}\mathbf{U} = \mathbf{I}
\end{eqnarray}
thus,
\begin{eqnarray}
\text{Tr}(\mathbf{I}) = \text{Tr}(\mathbf{U}^{\dagger}\mathbf{U}) = \sum_{jk}U_{jk}U^{\dagger}_{kj} = \sum_{jk}U_{jk}U^{*}_{jk}
\end{eqnarray}
Here, $\text{Tr}(\mathbf{I})$ equals to the number of electronic states. Similarly, 
\begin{eqnarray}
\sum_{jk}[\mathbf{U}^2]_{jk}[\mathbf{U}^{*}]^2_{jk} = \text{Tr}(\mathbf{I})
\end{eqnarray}
Hence, Eq.~({\ref{eq:traceofTsquared}}) becomes
\begin{eqnarray}
\text{Tr}(|\log(\mathbf{U'})|^2)&\approx&\frac{3}{4}\text{Tr}((\mathbf{U}^{*})^2) + \frac{3}{4}\text{Tr}(\mathbf{U}^2) - 4(\text{Tr}(\mathbf{U}) + \text{Tr}(\mathbf{U}^{*})) + \frac{13}{2} \text{Tr}(\mathbf{I})\nonumber\\
&=& Re\Big(\frac{3}{2}\text{Tr}(\mathbf{U}^2) - 8\text{Tr}(\mathbf{U})\Big) + \frac{13}{2}\text{Tr}(\mathbf{I})
\label{eq:poly_minimize}
\end{eqnarray}
Since we want to minimize this function by changing the phase of each column of $\mathbf{U}$, we can drop the constant term.

Thus, in the end, up to second order, the overlap matrix $\mathbf{U}$ should satisfy the following two conditions:
\begin{itemize}
    \item $\det(\mathbf{U}) = 1$
    \item $\text{Re}\big(\text{Tr}(3\mathbf{U}^2 - 16\mathbf{U})\big)$ is minimized
\end{itemize}

\subsection{The Two Algorithms for Real and Complex Regimes\label{subsec:both_ansatz}}
For completeness, we will now present a step-by-step algorithm for picking the phases of $\mathbf{U}$ in both real and complex regimes.
\subsubsection{Real Regime \label{subsec:rlh}}
For an $N$-state} problem, we will find the optimal combination of signs between columns of $\mathbf{U}$ by minimizing $\text{Tr}(3\mathbf{U}^2 - 16\mathbf{U})$ through Jacobi sweeps\cite{edmiston1963localized} while maintaining $\det(\mathbf{U}) = +1$. Specifically, a flowchart is:
\begin{steps}
 \item For a real overlap matrix $\mathbf{U}$, the determinant must obey $\det(\mathbf{U}) = \pm1$. If $\det(\mathbf{U})=-1$, we change the sign of the first eigenvector $\ket{\phi^{ad}_{1}(\vec{R}(t+dt_c))}$.
 \item To maintain $\det(\mathbf{U})=+1$, we check whether we should simultaneously flip the signs of a pair of columns of $\mathbf{U}$. There are $N(N-1)/2$ pairs of indices for an N-state system. For each pair $j$, $k$, we minimize $\text{Tr}(3\mathbf{U}^2 - 16\mathbf{U})$ by calculating the difference $\Delta_{\rm{real}}^{jk}$ (see Appendix for the derivation of $\Delta_{\rm{real}}^{jk}$) as follows:
\begin{lstlisting}[escapeinside={\%*}{*)}]
  set $\text{flagc}$ = 1
  loop j = 1 : N
    loop k = (j + 1) : N
       $\Delta_{\rm{real}}^{jk}\equiv 3(U_{jj}^2+ U_{kk}^2) + 6(U_{jk}U_{kj})$
            $ + 8(U_{jj} + U_{kk})-\displaystyle\sum_{l}3(U_{jl}U_{lj} + U_{kl}U_{lk})$
       if $\Delta_{\rm{real}}^{jk}<0$%*\begin{eqnarray}
            \ket{\phi^{ad}_{j}(\vec{R}(t+dt_c))} = -\ket{\phi^{ad}_{j}(\vec{R}(t+dt_c))} \nonumber
            \end{eqnarray}
            \begin{eqnarray}
            \ket{\phi^{ad}_{k}(\vec{R}(t+dt_c))} = -\ket{\phi^{ad}_{k}(\vec{R}(t+dt_c))}
            \label{eq: real_sweep}
            \end{eqnarray}*)                set $\text{flagc}$ = 0
    end
  end
\end{lstlisting}
 \item If $\text{flagc} == 0$, return to Step 2.
\end{steps}
\subsubsection{Complex Regime\label{subsec:cph}}
To extend the ansatz above into the complex regime, there are two major differences. First, strictly speaking a complex logarithm function is multi-valued, and one might worry about whether our approach is even well-defined. Nevertheless, for our purposes (i.e. minimization), we need only construct $\text{Tr}(|\log(\mathbf{U})|^2) = \sum_{j = 1}^{N}|\delta_{j}|^2$ in Eqs.~(\ref{eq: decomposition}) and (\ref{eq: mat_diag}), and the principal value of a complex logarithm is always well-defined. In other words, we need only insist that $\forall j$, $\delta_{j} \in (-\pi, \pi]$, which should solve this first problem. 

A second, more important difference is that whereas the phases of $\mathbf{U}$ are arbitrary up to a $+/-$ sign in the real regime, in the complex regime each adiabatic state can carry a complex phase $\exp{(i\theta_{j})}$ for each adiabatic state $\ket{\phi^{ad}_{j}(\vec{R}(t+dt_c))}$. If we want to explore changing the relative phases of two states $j$ and $k$, while maintaining $\det(\mathbf{U}) = 1$, we will need to sweep over the following phase possibilities:
\begin{eqnarray}
\ket{\phi^{ad}_{j}(\vec{R}(t+dt_c))} &\rightarrow& \exp(i\theta_{jk})\ket{\phi^{ad}_{j}(\vec{R}(t+dt_c))}\nonumber\\ 
\ket{\phi^{ad}_{k}(\vec{R}(t+dt_c))} &\rightarrow& \exp(-i\theta_{jk})\ket{\phi^{ad}_{k}(\vec{R}(t+dt_c))}.
\end{eqnarray}
Compare with Eq.~(\ref{eq: real_sweep}) above. The final flowchart is then as follows:
\begin{steps}
 \item We must start with a reasonable initial guess for the phases of $\mathbf{U}$. For each column $l$, we search over all rows indexed by $m$, and we find the matrix element $U_{ml}$ with the greatest absolute value $[\text{abs}(U_{ml})]$ (most often, as from parallel transport, we will find $m=l$); we then insist that $U_{ml}$ should be real and positive by multiplying the whole column $l$ (i.e. the eigenvector $\ket{\phi^{ad}_{l}(\vec{R}(t+dt_c))}$) by the complex conjugate of its phase $\frac{\text{conj}(U_{ml})}{\text{abs}(U_{ml})}$. This is the ansatz of standard parallel transport.
 \item As a complex overlap matrix, $\det(\mathbf{U})$ can be complex. If $\det(\mathbf{U})=\exp{(i\alpha)}$, we change the phase of the first eigenvector
 \begin{eqnarray}
 \ket{\phi^{ad}_{1}(\vec{R}(t+dt_c))} = \exp{(-i\alpha)}  \ket{\phi^{ad}_{1}(\vec{R}(t+dt_c))}.
 \end{eqnarray}
 By changing the phase of the first column of $\mathbf{U}$, we now have $\det(\mathbf{U}) = +1$
 \item To maintain $\det(\mathbf{U})=+1$, we change the phases of a pair of columns of $\mathbf{U}$ simultaneously. There are $N(N-1)/2$ pairs of indices for an $N$-state system. For each pair $j$, $k$, we multiply eigenvector $\ket{\phi^{ad}_{j}(R(t+dt_c))}$ by $\exp{(i\theta_{jk})}$ and multiply eigenvector $\ket{\phi^{ad}_{k}(R(t+dt_c))}$ by $\exp{(-i\theta_{jk})}$; we choose $\theta_{jk}$ such that we minimize $\text{Re}\bigl(\text{Tr}(3\mathbf{U}^2 - 16\mathbf{U})\bigr)$ (see Eq.~(\ref{eq:poly_minimize})). The Jacobi sweeps are performed as follows:
\begin{lstlisting}[escapeinside={\%*}{*)}]
  Set $\text{flagc}$ = 1
  loop j = 1 : N
    loop k = (j + 1) : N
      %%Calculate the following four intermediate quantities
      $\Gamma_{1}^{jk}= \Bigl[\displaystyle\sum_{l}6\rm{Re}\Bigl(U_{lj} U_{jl} + U_{lk} U_{kl} \Bigr)\Bigr] - 12\rm{Re}(U_{jk} U_{kj}) - 6\rm{Re}(U_{jj}^2 + U_{kk}^2) - 16\rm{Re}(U_{jj}+U_{kk})$
      $\Gamma_{2}^{jk}= 3\rm{Re}(U_{jj}^2 + U_{kk}^2)$
      $\Xi_{1}^{jk}= \Bigl[\displaystyle\sum_{l}6\rm{Im}\Bigl(U_{lk} U_{kl} - U_{lj} U_{jl} \Bigr)\Bigr] - 6\rm{Im}(U_{kk}^2 - U_{jj}^2) - 16\rm{Im}(U_{kk} - U_{jj})$
      $\Xi_{2}^{jk}= 3\rm{Im}(U_{kk}^2 - U_{jj}^2) $
      $\Delta_{\rm{complex}}^{jk} = \Gamma_{1}^{jk}\cos(\theta_{jk}) + \Gamma_{2}^{jk} \cos(2\theta_{jk}) + \Xi_{1}^{jk}\sin(\theta_{jk})+ \Xi_{2}^{jk}\sin(2\theta_{jk})$
      %%Calculate $\theta_{jk}$ by the four intermediate quantities
      $\theta_{jk} = \theta_{jk}(\Gamma_{1}^{jk}, \Gamma_{2}^{jk}, \Xi_{1}^{jk}, \Xi_{2}^{jk})$
      %%Change the phases of state $j$ and $k$      %*\begin{eqnarray}
      \ket{\phi^{ad}_{j}(\vec{R}(t+dt_c))} = \exp(i\theta_{jk})\ket{\phi^{ad}_{j}(\vec{R}(t+dt_c))}
      \label{eq: complexsweep}
      \end{eqnarray}
      \begin{eqnarray}
      \ket{\phi^{ad}_{k}(\vec{R}(t+dt_c))} = \exp(-i\theta_{jk})\ket{\phi^{ad}_{k}(\vec{R}(t+dt_c))}
      \nonumber
      \end{eqnarray}*)
      if $\theta_{jk}\neq0$
        set $\text{flagc}$ = 0
    end
  end
\end{lstlisting}

\item If $\text{flagc}==0$, return to Step 3.
\end{steps}
The derivation of $\Delta_{\rm{complex}}^{jk}$ is in Appendix~\ref{apdx: derivation_rc}. To solve for $\theta_{jk}$, see Appendix~\ref{apdx: mindeltajk} and~\ref{apdx: cp_mat}.

In the end, we have calculated the overlap matrix $\mathbf{U}$ with the new eigenvectors having well-defined phases. (Therefore, if needed, we can numerically compute the logarithm of $\mathbf{U}$ by Schur decomposition (see Eq.~(\ref{eq:TlogU})).\cite{loring2014computing})
\subsection{An approximate solution based on parallel transport\label{subsec: approx_method}}
For most dynamical calculations, choosing the signs of adiabatic states needs to be very fast: it should be \textit{much} faster, for instance, than diagonalization itself. Now, the protocol above suggests looping over all pairs of adiabatic states and performing Jacobi sweeps until convergence is attained (and the signs are fixed). In practice, for a very large matrix, this scheme could take time -- although so far, in our experience, the protocol above is always much faster than diagonalizing the electronic Hamiltonians. 

Nevertheless, in practice, one might want to fix some adiabatic state phases using parallel transport (i.e. set $U_{jj}\approx 1$ for some $j$) and then pick other adiabatic state phases with the more advanced scheme above. For this purpose, one would need a cutoff. In our experience, a natural cutoff should be $U_{jj}\leq1-\frac{2}{N}$. A great deal of numerical investigation suggests that if $|U_{jj}| > 1-\frac{2}{N}, \forall j$, then parallel transport (i.e. setting $U_{jj}$ positive) already minimizes $\text{Tr}(|\log(\mathbf{U})|^2)$. If some $U_{jj}$ satisfy $|U_{jj}| > 1-\frac{2}{N}$ (Class I) and other $U_{jj}$ satisfy $|U_{jj}| < 1-\frac{2}{N}$ (Class II), there is no guarantee that parallel transport is good enough for any state. Nevertheless, if the computation demands are heavy enough, we would recommend setting $U_{jj}$ to be real and positive for states in Class I and deciding $U_{jj}$ with Jacobi sweeps for Class II states. This approximation can be implemented in both the real and the complex regimes.

\section{Results \label{sec:3}}
\subsection{Numerical Test on the algorithm in the real regime}
Before applying the algorithm to a real dynamical model problem, we will generate a set of real random orthogonal matrices $\{\mathbf{U}\}$,\cite{diaconis1987subgroup} and test if our algorithm can find the optimal rotation matrix by optimizing the signs of all of the columns of $\mathbf{U}$. In principle, we might not be able to find the proper rotation matrix with the smallest $\text{Tr}(|\log(\mathbf{U})|^2)$ by truncating the Taylor series at second order {in $\mathbf{U}$} (see Section~\ref{subsubsec: min_poly_U}). Rather, we should really calculate the direct target function $\text{Tr}(|\log(\mathbf{U})|^2)$ explicitly (which is an infinite order Taylor series {in $\mathbf{U}$}). However, to show statistically that our algorithm works well enough, for a set of 1000 random unitary matrices of dimension $N$ in Table \ref{table: results}, {we will assess how well our algorithm indeed recovers the optimal matrix with the smallest $\text{Tr}(|\log(\mathbf{U})|^2)$. 
Note further that, in principle, the algorithm in Eqs.~({\ref{eq: real_sweep}}) and~({\ref{eq: complexsweep}}) could fail also by finding a local minimum (as opposed to a global minimum) of $\rm{Re}(\text{Tr}(3\mathbf{U}^2-16\mathbf{U}))$, and so to assess our approach, we will also benchmark how well Jacobi sweeps find the global minimum of our target function $\rm{Re}(\text{Tr}(3\mathbf{U}^2-16\mathbf{U}))$.
As a }side note, in all cases, our method is able to find a proper rotation matrix $\mathbf{U}$ with a real matrix logarithm, i.e. $\mathbf{U}$ has no eigenvalues equal to $-1$.
\begin{table}[]
\begin{tabular}{|c|c|c|c|c|}
\hline
\begin{tabular}[c]{@{}c@{}}Matrix \\Dimension \\N\end{tabular} & \begin{tabular}[c]{@{}c@{}}Number of \\matrices locating\\ the global\\ minimum of\\ $\text{Tr}(3\mathbf{U}^2 - 16\mathbf{U})$ \end{tabular} &
\begin{tabular}[c]{@{}c@{}}$\langle \text{Tr}(3\mathbf{U}^2 - 16\mathbf{U})\rangle/13N-$\\  $\langle \text{Tr}(3\mathbf{U}_{\text{global}}^2 - 16\mathbf{U}_{\text{global}})\rangle/13N$
\end{tabular} &\begin{tabular}[c]{@{}c@{}}Number of \\matrices locating\\ the global\\ minimum of \\ $\text{Tr}(|\log(\mathbf{U})|^2)$\end{tabular} &
\begin{tabular}[c]{@{}c@{}}$\langle\text{Tr}(|\log(\mathbf{U})|^2)\rangle/N$ \\ $-\langle \text{Tr}(|\log(\mathbf{U}_{\text{global}})|^2)\rangle/N$ \end{tabular}
\\
\hline
2 & 1000 & 0& 1000 & 0 \\ \hline
3 & 1000 & 0 & 1000 & 0 \\ \hline
4 & 997 & 0.106 & 980 & 0.075 \\ \hline
5 & 995 & 0.0726 & 975 & 0.063 \\\hline
6 & 988 & 0.0367 & 950 & 0.04 \\\hline
8 & 985 & 0.0251 & 865 & 0.0673 \\\hline
10 & 982 & 0.0114 & 720 & 0.0761 \\\hline
\end{tabular}
\caption{Results of a simple test of our algorithm with 1000 random unitary matrices. Note that for all cases and all dynamics, our algorithm finds a real proper rotation matrix $\mathbf{U}$ with a real matrix logarithm (i.e. $\mathbf{U}$ has no eigenvalues equal to $-1$). {However, our algorithm does not necessarily locate the globally optimal $\mathbf{U}$ for either the direct target function $\text{Tr}(3\mathbf{U}^2 - 16\mathbf{U})$ or the indirect target function $\text{Tr}(|\log(\mathbf{U})|^2)$, especially when $N$ grows larger. Nevertheless, Jacobi sweeps fail for only $18/1000$ test cases at finding the global minimum of $\text{Tr}(3\mathbf{U}^2 - 16\mathbf{U})$, and the total deviation $(\langle \text{Tr}(3\mathbf{U}^2 - 16\mathbf{U})\rangle - \langle \text{Tr}(3\mathbf{U}_{\text{global}}^2 - 16\mathbf{U}_{\text{global}})\rangle)/13N$ above the optimal $\mathbf{U}_{\text{global}}$} is very small in all cases, suggesting that the Jacobi sweeps algorithm is fairly robust. Admittedly, the failure rate for finding the true global minimum $\mathbf{U}$ is higher for the $\text{Tr}(|\log(\mathbf{U})|^2)$ criterion, which is a clear indication of the shortcomings of approximating a matrix logarithm with a polynomial. Nevertheless, as shown in the last column, the matrices $\mathbf{U}$ as obtained from our algorithm do not have very large deviations and we believe that they should be good enough for dynamics. See the results below and Appendix~\ref{apdx: failtests}.}
    \label{table: results}
\end{table}

{In Table~\ref{table: results}, we benchmark our algorithm as a function of vector space dimension $N$. As the dimension $N$ grows larger, we find that it does become more difficult to locate the global minimum of $\mathbf{U}$ for either the direct or indirect target functions. For $N = 10$, our algorithm fails to find the global minimum of $\rm{Re}(\text{Tr}(3\mathbf{U}^2-16\mathbf{U}))$ with probability 1.8\% and we fail to find the global minimum of $\text{Tr}(|\log(\mathbf{U})|^2)$ with the approximate polynomial with probability 28\%. Nevertheless, in all cases, Table~\ref{table: results} also demonstrates that the $\mathbf{U}$ found by our algorithm is {probably} good} enough. For instance, the deviation $\text{Tr}(|\log(\mathbf{U})|^2)-\text{Tr}(|\log(\mathbf{U}_{\text{global}})|^2)/N$ is never very large after we perform a Jacobi sweeps minimization: the deviation is only 0.0761 for the case of a $10\times10$ matrix. Overall, our belief is that the algorithm above should perform very well in practice.
Although we have not rigorously tested how large a deviation we can tolerate for accurate dynamics (but see Section~\ref{subsec: dynamicsmodel}), all data so far indicates that, if there are nearly equivalent sign conventions with small $\text{Tr}(|\log(\mathbf{U})|^2)$ or $\text{Tr}(3\mathbf{U}^2 - 16\mathbf{U})$, the exact choice of sign will not have large consequences; however, large  dynamical errors will arise if we select a sign convention that is not one of the nearly equivalent minima (and with a signficantly larger value of $\text{Tr}(|\log(\mathbf{U})|^2)$.  For the algorithm presented above, we find that small deviations from the target function global minima will arise only when $\mathbf{U}$ is large and very dense (with few zeros), but these are not expected to be common situations with reasonable simulation time steps. For details, see Appendix~\ref{apdx: failtests}.

\subsection{Testing the algorithm with a simple model problem in the complex regime\label{subsec: dynamicsmodel}}
In a companion paper, we tested Floquet Fewest Switch Surface Hopping (F-FSSH) on a few model problems with real Hamiltonians.\cite{zeyu2019nonadiabatic} In that paper, we used the algorithm above to compute $\mathbf{U}$ in the context of real Hamiltonians. In this paper, we will focus on a similar F-FSSH model problem but now with complex diabatic couplings in the Floquet picture, so that we can test the algorithm for complex Hamiltonians.

With this goal in mind, consider Tully's simple avoided crossing model problem modified to be time-dependent as the follows:
\begin{eqnarray}
H^{el}_{00}(R) &=& A[1-\exp(-B\times R)], \quad  R>0,\nonumber \\
\label{eq:T1st}
H^{el}_{00}(R) &=& - A[1-\exp(B\times R)], \quad  R<0, \\\nonumber
H^{el}_{11}(R) &=& -H^{el}_{00}(R), \\ \nonumber
H^{el}_{10}(R, t) &=& H^{el}_{01}(R, t) = C\exp(-D\times R^2)\cos(\omega t+\zeta).
\end{eqnarray}
Unless stated otherwise, all parameters will be chosen the same as in Ref.~[\onlinecite{zeyu2019nonadiabatic}], $A = 0.01$, $B = 1.6$, $C = 0.005$, $D = 1.0$, $\omega = 0.012$ and we set $dt = 1$. In Ref.~[\onlinecite{zeyu2019nonadiabatic}], we set $\zeta = 0$ such that the only periodic function was the cosine (which has real Fourier components). When an arbitrary phase $\zeta$ is introduced, however,
\begin{eqnarray}
H^{el}_{10}(R, t) = H^{el}_{01}(R, t) = C\exp(-D\times R^2)\cos(\omega t + \zeta) = V(R)\cos(\omega t + \zeta).
\end{eqnarray}
Thus, the Floquet Hamiltonian after the Fourier-type transformation becomes complex although the electronic Hamiltonian is still real.
\begin{eqnarray}
\cos(\omega t + \zeta) = \frac{\cos(\zeta) + i\sin(\zeta)}{2}\exp({i\omega t}) +\frac{\cos(\zeta) - i\sin(\zeta)}{2}\exp({-i\omega t})
\end{eqnarray}

Of course, a smart solution would be to shift the time coordinate with $t'=t+\zeta/\omega$ and change $t$ to $t'$ accordingly in Eqs. (45) and (46) (in Ref. [\onlinecite{zeyu2019nonadiabatic}]). Nevertheless, for our purposes, benchmarking F-FSSH with a trivially complex Floquet Hamiltonian will be a straightforward test of our algorithm for complex Hamiltonians above. Note that if we were to invoke a non-trivially complex Hamiltonian, we would need to discuss Berry's force and an approximate hopping direction, \cite{miao2019extension, subotnik2019demonstration} which would only complicate the present paper (and will be addressed in a future publication). Note that, when running F-FSSH, one would find that, unless we implement a robust ansatz in the complex regime, multiple trivial crossings at the origin can result in transitions to the wrong dressed states. 

\begin{figure}
    \centering
    \includegraphics[width=\textwidth]{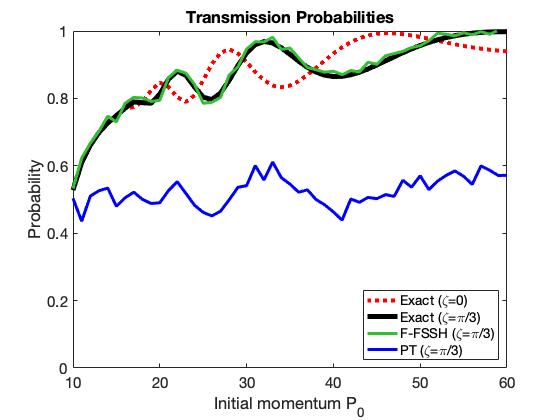}
    \caption{Transmission Probabilities on diabat $\ket{0}$ for modified complex simple avoided crossing problem. The black line shows the exact results with phase $\zeta = \frac{\pi}{3}$. The red dotted line represents the exact results with $\zeta = 0$. When $\zeta \neq 0$, there is a clear shift of the phase of oscillation. The green line plots "F-FSSH ($\zeta = \frac{\pi}{3}$)" data. Note that F-FSSH predicts accurately the transmission probabilities using the sign convention outlined above. However, as shown by the blue line labeled "PT ($\zeta = \frac{\pi}{3}$)", {straightforward parallel transport results demonstrate that one can find results that are very incorrect if one uses a different convention (where one forces $U_{jj}$ to be real and positive even if $U_{jj}$ is small). }As is well known, choosing the phases of adiabatic states is critically important here.{\cite{akimov2018simple}}}
    \label{fig: FSH_cpdc}
\end{figure}

Exact and F-FSSH results for $\zeta = \pi / 3$ are shown in Fig.~{\ref{fig: FSH_cpdc}}, in comparison to the results with $\zeta = 0$ (red dotted line, which is the same as the black line in Fig. 2(f) in Ref.~[\onlinecite{zeyu2019nonadiabatic}]). Apparently, changing $\zeta$ leads to a shift of the oscillation on the exact black line. As in Ref.~[\onlinecite{zeyu2019nonadiabatic}], F-FSSH performs well {if we adopt our phase convention, but the algorithm fails completely if we use simple parallel transport (where we force $U_{jj}$ to be real and positive no matter how small $U_{jj}$ is) and a reasonable time step.} These results confirm that in practice, choosing the phases of $\mathbf{U}$ following the algorithm in Section~\ref{subsec:cph} is robust, efficient, and essential for accuracy. Note also that we recover exactly the same dynamics if we use the approximate scheme for $\mathbf{U}$, whereby only some eigenvector signs are optimized beyond parallel transport (see Section~\ref{subsec: approx_method}).
\section{Conclusions \label{sec:4}}
In conclusion, we have presented an ansatz that can efficiently and universally pick phases for parametrized sets of eigenvectors so as to evaluate the time-derivative coupling matrix $\mathbf{T}$ as smoothly as possible for both avoided crossings and trivial crossings {and we have tested our ansatz for both real and complex model problems that contain multiple pair-wise trivial crossings at the same time step. To minimize $\text{Tr}(|\log(\mathbf{U})|^2)$, we make the ansatz that we must enforce (i) the exact constraint that $\det(\mathbf{U}) = +1$ and (ii) the approximate constraint that $\rm{Re}(\text{Tr}(3\mathbf{U}^2-16\mathbf{U}))$ should be minimized. Despite any limitations from the uncontrolled quadratic Taylor series approximation of $\text{Tr}(|\log(\mathbf{U})|^2)$, all results confirm that the ansatz is robust and efficient. In general, our constraints yield a $\mathbf{U}$ that has very small value of $\text{Tr}(|\log(\mathbf{U})|^2)$, and very often we reach the global minimum. Overall, we are quite }confident that our ansatz should be very powerful as far as calculating time-derivative couplings or non-adiabatic couplings for surface hopping calculations or other non-adiabatic dynamics formalisms.

As far as performance, the Jacobi sweeps in Section~\ref{subsec:both_ansatz} (in both the real and complex regimes) converge fast so that there is no significant additional computational cost to implementing our ansatz in surface hopping; one can even apply Jacobi sweeps to a subgroup of the set of adiabatic states and achieve effectively the same results with basically zero cost (see Section~\ref{subsec: approx_method}). {Alternatively, in the future, one can also imagine running Monte Carlo to minimize $\rm{Re}(\text{Tr}(3\mathbf{U}^2-16\mathbf{U}))$, rather than performing Jacobi sweeps (which is basically steepest descent and is usually not optimal for large systems\footnote{Monte Carlo should also limit any propensity to find local minima for $\rm{Re}(\text{Tr}(3\mathbf{U}^2-16\mathbf{U}))$ rather than the global minimum. Admittedly, however, Monte Carlo cannot solve the problem of a inadequate Taylor series, or help us minimize $\text{Tr}(|\log(\mathbf{U})|^2)$ globally if $\text{Tr}(|\log(\mathbf{U})|^2)$ is not minimized by $\rm{Re}(\text{Tr}(3\mathbf{U}^2-16\mathbf{U}))$}). }In practice, running nuclear dynamics and evaluating electronic structure will remain the only bottleneck in dynamics calculations. In the end, the present algorithm, combined with the methods of Ref.~[\onlinecite{jain2016efficient}] should allow us to run FSSH in a blackbox manner with reasonably large time steps, and achieve real gains in cost, while never worrying about trivial or non-trivial crossings.
\begin{acknowledgments}
This work was supported by the U.S. Air Force Office of Scientific
Research (USAFOSR) AFOSR Grants No. FA9550-18-1-0497 and FA9550-18-1-0420. 
T. Q. acknowledges the support from the Vagelos Institute for Energy Science and Technology (VIEST).
A.M.R acknowledges the support of the US Department of Energy, Office of Basic Energy Sciences, under grant DE-SC0019281.
\end{acknowledgments}
\appendix

\section{Derivation of $\Delta_{\rm{real}}^{jk}, \Delta_{\rm{complex}}^{jk}, \Gamma_{1}^{jk}, \Gamma_{2}^{jk}, \Xi_{1}^{jk}, \Xi_{2}^{jk}$ in Section~\ref{subsec:both_ansatz} \label{apdx: derivation_rc}}
Let us calculate the difference in $\text{Re}\big(\text{Tr}(3\mathbf{U}^2 - 16\mathbf{U})\big)$ between overlap matrices $\mathbf{U}$ before and after we multiply state $j$ by $\exp{(i\theta_{jk})}$ and state $k$ by $\exp{(-i\theta_{jk})}$. We will start with $\Delta_{\rm{complex}}^{jk}$ and then $\Delta_{\rm{real}}^{jk}$ can be obtained naturally by assigning $\theta_{jk} = \pi$.
We denote the original overlap matrix as $\mathbf{U}$ and the overlap matrix with the phases of two column changed as $\mathbf{U}'$, i.e.
\begin{eqnarray}
U'_{mn}  = 
\begin{cases}
U_{mn} & \text{if } n\neq j, k, \\
U_{mn}\exp(i\theta_{jk}) & \text{if } n=j, \\
U_{mn}\exp(-i\theta_{jk}) & \text{if } n=k.
\end{cases}
\end{eqnarray}
Then,
\begin{eqnarray}
\text{Tr}(\mathbf{U}'-\mathbf{U}) = U_{jj}\exp(i\theta_{jk}) + U_{kk}\exp(-i\theta_{jk}) - U_{jj} - U_{kk}.
\end{eqnarray}
\begin{eqnarray}
\text{Tr}(\mathbf{U}'^2 - \mathbf{U}^2) &=& \sum_{l}2U_{jl}U_{lj}\big(\exp(i\theta_{jk}) - 1\big) + 2U_{kl}U_{lk}\big(\exp(-i\theta_{jk}) - 1\big) \nonumber\\
&&- 2U_{jk}U_{kj}\big(\exp(i\theta_{jk}) + \exp(-i\theta_{jk}) - 2\big) + U_{jj}^2\big(\exp(i\theta_{jk}) -1\big)^2\\\nonumber
&&+ U_{kk}^2\big(\exp(-i\theta_{jk}) -1\big)^2.
\end{eqnarray}
Next, we take the real part of each difference:
\begin{eqnarray}
\rm{Re}(\text{Tr}(\mathbf{U}' - \mathbf{U})) = \rm{Re}(U_{jj} + U_{kk})\cos\theta_{jk} + \rm{Im}(U_{kk} - U_{jj})\sin\theta_{jk}- U_{jj} - U_{kk}.
\label{trace_dif_fs}
\end{eqnarray}
\begin{eqnarray}
\rm{Re}(\text{Tr}(\mathbf{U}'^2-\mathbf{U}^2)) &=& \sum_{l}\{2\rm{Re}(U_{jl}U_{lj} + U_{kl}U_{lk})(\cos\theta_{jk}-1) + 2\rm{Im}(U_{kl}U_{lk}-U_{jl}U_{lj})\sin\theta_{jk}\} 
\nonumber\\
- 4&\text{Re}&(U_{jk}U_{kj})(\cos\theta_{jk} - 1) + \rm{Re}(U_{jj}^2+ U_{kk}^2)(\cos(2\theta_{jk}) - 2\cos\theta_{jk} + 1)
\label{trace_dif_sq}
\\\nonumber 
&&+ \rm{Im}(U_{kk}^2-U_{jj}^2)(\sin(2\theta_{jk}) - 2\sin\theta_{jk}).
\end{eqnarray}
Since we want to minimize $\Delta_{\rm{complex}}^{jk} \sim \text{Re}\big(\text{Tr}(3\mathbf{U}'^2 - 16\mathbf{U}')\big)$, we want the difference $\text{Re}\big(\text{Tr}\big(3(\mathbf{U}'^2 - \mathbf{U}^2) - 16(\mathbf{U}' - \mathbf{U})\big)\big)$ to be minimized, and so all constants for a given $\mathbf{U}$ can be dropped. By combining the coefficients of the cosine or sine functions, we obtain
\begin{eqnarray}
&\Delta&_{\rm{complex}}^{jk}\equiv 3\rm{Re}(U_{jj}^2+ U_{kk}^2)\cos(2\theta_{jk}) + 3\rm{Im}(U_{kk}^2-U_{jj}^2)\sin(2\theta_{jk})  \nonumber\\\nonumber
&+&\Big\{\big(\sum_{l}6\rm{Re}(U_{jl}U_{lj} + U_{kl}U_{lk})\big) - 12\rm{Re}(U_{jk}U_{kj}) - 6\rm{Re}(U_{jj}^2+ U_{kk}^2) - 16\rm{Re}(U_{jj} + U_{kk})\Big\}\cos\theta_{jk} \\
&+& \Big\{\big(\sum_{l}6\rm{Im}(U_{kl}U_{lk}-U_{jl}U_{lj})\big)-6\rm{Im}(U_{kk}^2-U_{jj}^2)-16\rm{Im}(U_{kk} - U_{jj})\Big\}\sin\theta_{jk} \nonumber\\
&\equiv&\Gamma_{2}^{jk}\cos(2\theta_{jk}) + \Xi_{2}^{jk}\sin(2\theta_{jk}) + \Gamma_{1}^{jk}\cos\theta_{jk} + \Xi_{1}^{jk}\sin\theta_{jk}
\label{eq: complex_dif}
\end{eqnarray}
Here, the coefficients are the same as those in Section~{\ref{subsec:cph}}.

For the real case, rather than minimizing a function, we would like simply to check whether a difference $\Delta_{\rm{real}}^{jk}$ is positive or negative. To that end, we will keep the constants that we dropped above between Eqs.~({\ref{trace_dif_fs}}-{\ref{trace_dif_sq}}) and Eq.~(\ref{eq: complex_dif}). By setting $\theta_{jk}=\pi$, we can obtain the relevant difference quickly by combining Eq.~({\ref{trace_dif_fs}}) and Eq.~({\ref{trace_dif_sq}}) in the real regime:
\begin{eqnarray}
\Delta_{\rm{real}}^{jk}\equiv 3(U_{jj}^2+ U_{kk}^2) + 6(U_{jk}U_{kj}) + 8(U_{jj} + U_{kk}) 
-\sum_{l}3(U_{jl}U_{lj} + U_{kl}U_{lk})
\end{eqnarray}

\section{Minimization of $\Delta_{\rm{complex}}^{jk}$\label{apdx: mindeltajk}}
In order to find $\theta_{jk}$ by minimizing $\Delta_{\rm{complex}}^{jk}$ (Eq.~(\ref{eq: complex_dif})),
\begin{eqnarray}
\Delta_{\rm{complex}}^{jk} \equiv \Gamma_{2}^{jk}\cos(2\theta_{jk}) + \Xi_{2}^{jk}\sin(2\theta_{jk}) + \Gamma_{1}^{jk}\cos\theta_{jk} + \Xi_{1}^{jk}\sin\theta_{jk}
\label{eq:cfunctomin}
\end{eqnarray}
we take the derivative with respect to $\theta_{jk}$
\begin{eqnarray}
\frac{d\Delta_{\rm{complex}}^{jk}}{d\theta_{jk}} \equiv 2\Xi_{2}^{jk}\cos(2\theta_{jk}) - 2\Gamma_{2}^{jk}\sin(2\theta_{jk}) - \Gamma_{1}^{jk}\sin\theta_{jk} + \Xi_{1}^{jk}\cos\theta_{jk}
\label{eq:firstderivative}
\end{eqnarray}
At a minimum, this derivative must equal 0.

To solve this equation, we set $x=\cos\theta_{jk}$. Eq.~({\ref{eq:firstderivative}}) becomes:
\begin{eqnarray}
2\Xi_{2}^{jk}(2x^2-1) - 4\Gamma_{2}^{jk}x\sqrt{1-x^2} - \Gamma_{1}^{jk}\sqrt{1-x^2} + \Xi_{1}^{jk}x=0
\end{eqnarray}
By rearranging the equation, we obtain a quartic equation:
\begin{eqnarray}
16[(\Xi_{2}^{jk})^2+(\Gamma_{2}^{jk})^2]x^4 &+& 8(\Xi_{1}^{jk}\Xi_{2}^{jk} + \Gamma_{1}^{jk}\Gamma_{2}^{jk}) x^3 + [(\Xi_{1}^{jk})^2 + (\Gamma_{1}^{jk})^2-16(\Xi_{2}^{jk})^2 - 16(\Gamma_{2}^{jk})^2]x^2\nonumber \\ &+&  4(\Xi_{1}^{jk}\Xi_{2}^{jk} + 2\Gamma_{1}^{jk}\Gamma_{2}^{jk})x + [4(\Xi_{2}^{jk})^2 - (\Gamma_{1}^{jk})^2] = 0
\label{eq:quarticeq}
\end{eqnarray}

There are three scenarios. First, when $(\Xi_{2}^{jk})^2+(\Gamma_{2}^{jk})^2 = 0$, Eq.~({\ref{eq:firstderivative}}) reduces to 
\begin{eqnarray}
\Gamma_{1}^{jk}\sin\theta_{jk} = \Xi_{1}^{jk}\cos\theta_{jk}
\end{eqnarray}
Note that $\theta_{jk}$ has a period $2\pi$ while $\text{arctan}(x)$ has a period $\pi$, and thus, there are two roots,
\begin{eqnarray}
\theta_{jk}^{1} = \text{arctan}\Big(\frac{\Xi_{1}^{jk}}{\Gamma_{1}^{jk}}\Big)
\end{eqnarray}
\begin{eqnarray}
\theta_{jk}^{2} = \text{arctan}\Big(\frac{\Xi_{1}^{jk}}{\Gamma_{1}^{jk}}\Big) + \pi
\end{eqnarray}
We pick the $\theta_{jk}$ which makes Eq.~({\ref{eq:cfunctomin}}) smallest. The physical meaning is that we have encountered a pair-wise trivial crossing between state $j$ and $k$.

The second scenario is that we have to solve this quartic equation without any possible reduction. We can either apply the general form of the solution to any quartic equation, or by constructing the companion matrix of Eq.~({\ref{eq:quarticeq}}) (See Appendix~\ref{apdx: cp_mat}).
For a quartic equation, we may have 4 real roots at most. Again, $\theta_{jk}$ has a period $2\pi$. Thus, for each real root, there are two possible $\theta_{jk}$.
\begin{eqnarray}
\theta_{jk}^{1} = \text{arccos}(x)
\end{eqnarray}
\begin{eqnarray}
\theta_{jk}^{2} = -\text{arccos}(x)
\end{eqnarray}
In total, we have 8 roots at most. Since we are trying to find the minimum of $\Delta_{\rm{complex}}^{jk}$, we may simply calculate $\Delta_{\rm{complex}}^{jk}$ for all real roots and choose the root with smallest $\Delta_{\rm{complex}}^{jk}$.

In principle, there is a third scenario that is the most tricky: all four coefficients could effectively be zeros. In this case, the physical meaning is that within the time step, we encounter one or more multi-state trivial crossing between state $j$, $k$ and at least one other state, and there are no other adiabatic states $\{m\}$ that talk to either $j$ or $k$: $\sum_{m}U_{jm}U_{mj} = \sum_{k}U_{jk}U_{kj} = 0$. For this situation, the ansatz above, based on truncation at second order, is no longer valid. In theory, one could derive a similar but higher-order algorithm to solve for the phases of $\mathbf{U}$ to minimize $\text{Tr}(|\log(\mathbf{U})|^2)$ more accurately. Nevertheless, we believe this scenario should not be very physically relevant, since the phases of different adiabatic states ($j$ and $k$) can matter only when two systems interact directly or indirectly and in such a situation, it seems very unlikely there will not be one single state that interacts with either $j$ or $k$ at the same time (so that $(U^2)_{jj} = (U^2)_{kk} = 0$). As a practical matter, we believe truncating at second order should be sufficient.

\section{Finding roots of a polynomial \label{apdx: cp_mat}}
For a monic polynomial equation with real coefficients
\begin{eqnarray}
p(x) = x^{n} + a_{n-1}x^{n-1} + \dots + a_{1}x + a_{0} = 0,
\end{eqnarray}
there will be n roots (and some of them may be complex). The roots can be obtained by constructing the Frobenius companion matrix $\mathbf{C}(p)$.
\[\mathbf{C}(p) = \left[
\begin{array}{c c c c c }
0 & 0 & \dots & 0 & -a_{0}\\
1 & 0 & \dots & 0 & -a_{1} \\
0 & 1 & \dots & 0 & -a_{2}\\
\vdots & \vdots & \ddots & \vdots & \vdots \\
0 & 0 & \dots & 1 &-a_{n-1} \\
\end{array} \right] \]
The eigenvalues which can be obtained by performing a Schur decomposition are the roots of the polynomial equation $p(x)$.

\section{The limitations of our algorithm\label{apdx: failtests}}
In this appendix, we want to explicitly explore the different cases where our algorithm breaks down. {We will consider explicitly three different $4\times4$ unitary matrices $\mathbf{A}$, $\mathbf{B}$ and $\mathbf{C}$.}

{The first matrix $\mathbf{A}$ has the following sign possibilities:
\[\mathbf{A}_{1} = \left[
\begin{array}{c c c c}
    0.6575 & -0.3565 & -0.6354 & -0.1920 \\
   0.1351 & 0.6081 & -0.4038 & 0.6700 \\
   0.0916 & 0.6991 & -0.0847 & -0.7041 \\
   0.7355 & 0.1199 & 0.6527 & 0.1363 \\
\end{array} \right] \]
\[\mathbf{A}_{2} = \left[
\begin{array}{c c c c}
    0.6575 & -0.3565 & 0.6354 & 0.1920 \\
   0.1351 & 0.6081 & 0.4038 & -0.6700 \\
   0.0916 & 0.6991 & 0.0847 & 0.7041 \\
   0.7355 & 0.1199 & -0.6527 & -0.1363 \\
\end{array} \right] \]
\[\mathbf{A}_{3} = \left[
\begin{array}{c c c c}
    0.6575 & 0.3565 & 0.6354 & -0.1920 \\
   0.1351 & -0.6081 & 0.4038 & 0.6700 \\
   0.0916 & -0.6991 & 0.0847 & -0.7041 \\
   0.7355 & -0.1199 & -0.6527 & 0.1363 \\
\end{array} \right] \]
\[\mathbf{A}_{4} = \left[
\begin{array}{c c c c}
    -0.6575 & -0.3565 & 0.6354 & -0.1920 \\
   -0.1351 & 0.6081 & 0.4038 & 0.6700 \\
   -0.0916 & 0.6991 & 0.0847 & -0.7041 \\
   -0.7355 & 0.1199 & -0.6527 & 0.1363 \\
\end{array} \right] \]
\[\mathbf{A}_{5} = \left[
\begin{array}{c c c c}
    -0.6575 & -0.3565 & -0.6354 & 0.1920 \\
   -0.1351 & 0.6081 & -0.4038 & -0.6700 \\
   -0.0916 & 0.6991 & -0.0847 & 0.7041 \\
   -0.7355 & 0.1199 & 0.6527 & -0.1363 \\
\end{array} \right] \]
\[\mathbf{A}_{6} = \left[
\begin{array}{c c c c}
    0.6575 & 0.3565 & -0.6354 & 0.1920 \\
   0.1351 & -0.6081 & -0.4038 & -0.6700 \\
   0.0916 & -0.6991 & -0.0847 & 0.7041 \\
   0.7355 & -0.1199 & 0.6527 & -0.1363 \\
\end{array} \right] \]
\[\mathbf{A}_{7} = \left[
\begin{array}{c c c c}
    -0.6575 & 0.3565 & 0.6354 & 0.1920 \\
   -0.1351 & -0.6081 & 0.4038 & -0.6700 \\
   -0.0916 & -0.6991 & 0.0847 & 0.7041 \\
   -0.7355 & -0.1199 & -0.6527 & -0.1363 \\
\end{array} \right] \]
\[\mathbf{A}_{8} = \left[
\begin{array}{c c c c}
    -0.6575 & 0.3565 & -0.6354 & -0.1920 \\
   -0.1351 & -0.6081 & -0.4038 & 0.6700 \\
   -0.0916 & -0.6991 & -0.0847 & -0.7041 \\
   -0.7355 & -0.1199 & 0.6527 & 0.1363 \\
\end{array} \right] \]}

{The next matrix $\mathbf{B}$ has the following sign possibilities:
\[\mathbf{B}_{1} = \left[
\begin{array}{c c c c}
0.5987 & 0.1138 & 0.5219 & 0.5969 \\
-0.5288 & 0.5520 & 0.6321 & -0.1274 \\
-0.5694 & -0.1139 & -0.2188 & 0.7842 \\
0.1942 & 0.8182 & -0.5294 & 0.1121 \\
\end{array} \right] \]
\[\mathbf{B}_{2} = \left[
\begin{array}{c c c c}
0.5987 & 0.1138 & -0.5219 & -0.5969 \\
-0.5288 & 0.5520 & -0.6321 & 0.1274 \\
-0.5694 & -0.1139 & 0.2188 & -0.7842 \\
0.1942 & 0.8182 & 0.5294 & -0.1121 \\
\end{array} \right] \]
\[\mathbf{B}_{3} = \left[
\begin{array}{c c c c}
0.5987 & -0.1138 & -0.5219 & 0.5969 \\
-0.5288 & -0.5520 & -0.6321 & -0.1274 \\
-0.5694 & 0.1139 & 0.2188 & 0.7842 \\
0.1942 & -0.8182 & 0.5294 & 0.1121 \\
\end{array} \right] \]
\[\mathbf{B}_{4} = \left[
\begin{array}{c c c c}
-0.5987 & 0.1138 & -0.5219 & 0.5969 \\
0.5288 & 0.5520 & -0.6321 & -0.1274 \\
0.5694 & -0.1139 & 0.2188 & 0.7842 \\
-0.1942 & 0.8182 & 0.5294 & 0.1121 \\
\end{array} \right] \]
\[\mathbf{B}_{5} = \left[
\begin{array}{c c c c}
-0.5987 & 0.1138 & 0.5219 & -0.5969 \\
0.5288 & 0.5520 & 0.6321 & 0.1274 \\
0.5694 & -0.1139 & -0.2188 & -0.7842 \\
-0.1942 & 0.8182 & -0.5294 & -0.1121 \\
\end{array} \right] \]
\[\mathbf{B}_{6} = \left[
\begin{array}{c c c c}
0.5987 & -0.1138 & 0.5219 & -0.5969 \\
-0.5288 & -0.5520 & 0.6321 & 0.1274 \\
-0.5694 & 0.1139 & -0.2188 & -0.7842 \\
0.1942 & -0.8182 & -0.5294 & -0.1121 \\
\end{array} \right] \]
\[\mathbf{B}_{7} = \left[
\begin{array}{c c c c}
-0.5987 & -0.1138 & -0.5219 & -0.5969 \\
0.5288 & -0.5520 & -0.6321 & 0.1274 \\
0.5694 & 0.1139 & 0.2188 & -0.7842 \\
-0.1942 & -0.8182 & 0.5294 & -0.1121 \\
\end{array} \right] \]
\[\mathbf{B}_{8} = \left[
\begin{array}{c c c c}
-0.5987 & -0.1138 & 0.5219 & 0.5969 \\
0.5288 & -0.5520 & 0.6321 & -0.1274 \\
0.5694 & 0.1139 & -0.2188 & 0.7842 \\
-0.1942 & -0.8182 & -0.5294 & 0.1121 \\
\end{array} \right] \]}

{The third matrix $\mathbf{C}$ has the following sign possibilities:
\[\mathbf{C}_{1} = \left[
\begin{array}{c c c c}
0.1451 & 0.6731 & -0.7116 & 0.1397  \\
-0.9396  & -0.0885  & -0.3019 & -0.1351 \\
0.1431 & 0.1265 & -0.0437 & -0.9806 \\
-0.2751  & 0.7233 & 0.6329 & 0.0249 
\end{array} \right] \]
\[\mathbf{C}_{2} = \left[
\begin{array}{c c c c}
-0.1451 & -0.6731 & 0.7116 & -0.1397  \\
0.9396  & 0.0885  & 0.3019 & 0.1351 \\
-0.1431 & -0.1265 & 0.0437 & 0.9806 \\
0.2751  & -0.7233 & -0.6329 & -0.0249 
\end{array} \right] \]
\[\mathbf{C}_{3} = \left[
\begin{array}{c c c c}
0.1451 & 0.6731 & 0.7116 & -0.1397  \\
-0.9396  & -0.0885  & 0.3019 & 0.1351 \\
0.1431 & 0.1265 & 0.0437 & 0.9806 \\
-0.2751  & 0.7233 & -0.6329 & -0.0249 
\end{array} \right] \]
\[\mathbf{C}_{4} = \left[
\begin{array}{c c c c}
-0.1451 & -0.6731 & -0.7116 & 0.1397  \\
0.9396  & 0.0885  & -0.3019 & -0.1351 \\
-0.1431 & -0.1265 & -0.0437 & -0.9806 \\
0.2751  & -0.7233 & 0.6329 & 0.0249 
\end{array} \right] \]
\[\mathbf{C}_{5} = \left[
\begin{array}{c c c c}
0.1451 & -0.6731 & 0.7116 & 0.1397  \\
-0.9396  & 0.0885  & 0.3019 & -0.1351 \\
0.1431 & -0.1265 & 0.0437 & -0.9806 \\
-0.2751  & -0.7233 & -0.6329  & 0.0249 
\end{array} \right] \]
\[\mathbf{C}_{6} = \left[
\begin{array}{c c c c}
0.1451 & -0.6731 & -0.7116 & -0.1397  \\
-0.9396  & 0.0885  & -0.3019 & 0.1351 \\
0.1431 & -0.1265 & -0.0437 & 0.9806 \\
-0.2751  & -0.7233 & 0.6329  & -0.0249 
\end{array} \right] \]
\[\mathbf{C}_{7} = \left[
\begin{array}{c c c c}
-0.1451 & 0.6731 & 0.7116 & 0.1397  \\
0.9396  & -0.0885  & 0.3019 & -0.1351 \\
-0.1431 & 0.1265 & 0.0437 & -0.9806 \\
0.2751  & 0.7233 & -0.6329  & 0.0249 
\end{array} \right] \]
\[\mathbf{C}_{8} = \left[
\begin{array}{c c c c}
-0.1451 & 0.6731 & -0.7116 & -0.1397  \\
0.9396  & -0.0885  & -0.3019 & 0.1351 \\
-0.1431 & 0.1265 & -0.0437 & 0.9806 \\
0.2751  & 0.7233 & 0.6329  & -0.0249 
\end{array} \right] \]}

{In Tables \ref{table: signconvention}, \ref{table: failsignconvention} and \ref{table: localmin}, we analyze the matrices $\mathbf{A}$, $\mathbf{B}$ and $\mathbf{C}$ together with different sign conventions. The fact that $\mathbf{A}$, $\mathbf{B}$ and $\mathbf{C}$ are dense $4\times4$ matrices implies that four states are crossing with each other strongly (which is uncommon in reality). We report }the two relevant quantities $\text{Tr}(3\mathbf{U}^2 - 16\mathbf{U})$ and $\text{Tr}(|\log(\mathbf{U})|^2)$ for all possible sign conventions.
\begin{table}[ht]
    \centering
    \begin{tabular}{|c|c|c|}
    \hline
    Sign Convention  &  $\text{Tr}(3\mathbf{A}^2 - 16\mathbf{A})$ & $\text{Tr}(|\log(\mathbf{A})|^2)$ \\
    \hline
    $\mathbf{A}_{1}$ & -24.0463 & 6.8250  \\\hline
    $\mathbf{A}_{2}$ & -17.5786 & 7.7976  \\\hline
    $\mathbf{A}_{3}$ & -1.4704 & 11.8361  \\\hline
    $\mathbf{A}_{4}$ & 5.4582 & 14.0330 \\\hline
    $\mathbf{A}_{5}$ & 7.1824 & 14.6045 \\\hline
    $\mathbf{A}_{6}$ & 10.9494 & 16.8259 \\\hline
    $\mathbf{A}_{7}$ & 18.1041 & 17.1368 \\\hline
    $\mathbf{A}_{8}$ & 21.2694 & 22.2017 \\\hline
    \end{tabular}
    \caption{{An example for which our Jacobi sweeps method successfully locates global minimum of $\text{Tr}(|\log(\mathbf{A})|^2)$. The matrix $\mathbf{A}_{1}$ is the global minimum of both quantities $\text{Tr}(|\log(\mathbf{A})|^2)$ and $\text{Tr}(3\mathbf{A}^2 - 16\mathbf{A})$}}
    \label{table: signconvention}
\end{table}

{In table~{\ref{table: signconvention}}, for the $\mathbf{A}$ matrix, we show that the two quantities $\text{Tr}(3\mathbf{A}^2 - 16\mathbf{A})$ and $\text{Tr}(|\log(\mathbf{A})|^2)$ share the same trends for different sign conventions: minimizing $\text{Tr}(3\mathbf{A}^2 - 16\mathbf{A})$ is consistent with minimizing $\text{Tr}(|\log(\mathbf{A})|^2)$.}
\begin{table}[ht]
    \centering
    \begin{tabular}{|c|c|c|}
    \hline
    Sign Convention  &  $\text{Tr}(3\mathbf{B}^2 - 16\mathbf{B})$ & $\text{Tr}(|\log(\mathbf{B})|^2)$ \\
    \hline
    $\mathbf{B}_{2}$  & -18.6547 & 7.5890 \\\hline
    $\mathbf{B}_{1}$  & -19.5302 & 7.7673 \\\hline
    $\mathbf{B}_{4}$ & -2.1964 & 11.6490 \\\hline
    $\mathbf{B}_{3}$  & 1.6531 & 12.6090 \\\hline
    $\mathbf{B}_{6}$  & 6.8980 & 14.4798 \\\hline
    $\mathbf{B}_{7}$ & 13.8778 & 16.0589 \\\hline
    $\mathbf{B}_{5}$ & 13.7363 & 19.3474 \\\hline
    $\mathbf{B}_{8}$ & 21.5821 & 21.2692 \\\hline
    \end{tabular}
    \caption{{An example which our Jacobi sweeps method fails to locate the global minimum of $\text{Tr}(|\log(\mathbf{B})|^2)$. The quantity $\text{Tr}(|\log(\mathbf{B})|^2)$ is minimized by sign convention $\mathbf{B_{2}}$, but the quantity $\text{Tr}(3\mathbf{B}^2 - 16\mathbf{B})$ is minimized by sign convention $\mathbf{B}_{1}$. The difference (or deviation}) arises from a failure of the polynomial approximation $\text{Tr}(3\mathbf{B}^2 - 16\mathbf{B})\sim \text{Tr}(|\log(\mathbf{B})|^2)$.}
    \label{table: failsignconvention}
\end{table}
{In table~{\ref{table: failsignconvention}}, however, we show that our method cannot find the global minimum of $\text{Tr}(|\log(\mathbf{B})|^2)$, because $\text{Tr}(3\mathbf{B}^2 - 16\mathbf{B})$ is not completely consistent with $\text{Tr}(|\log(\mathbf{B})|^2)$. Nevertheless, the difference is small and our method does locate the second best sign convention as far as $\text{Tr}(|\log(\mathbf{B})|^2)$ is concerned.}
\begin{table}[ht]
    \centering
    \begin{tabular}{|c|c|c|}
    \hline
    Sign Convention &  $\text{Tr}(3\mathbf{C}^2 - 16\mathbf{C})$ & $\text{Tr}(|\log(\mathbf{C})|^2)$ \\
    \hline
    $\mathbf{C}_{1}$ & -9.686  & 10.2505 \\ \hline
    $\mathbf{C}_{2}$ & -8.4764 & 10.5041 \\ \hline
    $\mathbf{C}_{3}$ & -6.9735 & 10.7389 \\ \hline
    $\mathbf{C}_{4}$ & -4.5607 & 11.2894 \\ \hline
    $\mathbf{C}_{5}$ & 3.515   & 13.2578 \\ \hline
    $\mathbf{C}_{6}$ & 4.2351  & 13.5732 \\ \hline
    $\mathbf{C}_{7}$ & 9.5151  & 15.8454 \\ \hline
    $\mathbf{C}_{8}$ & 13.1854 & 18.9589 \\ \hline
    \end{tabular}
    \caption{{An example which our Jacobi sweeps method fails to locate the global minimum of either $\text{Tr}(3\mathbf{C}^2 - 16\mathbf{C})$ or $\text{Tr}(|\log(\mathbf{C})|^2)$. Both quantities $\text{Tr}(3\mathbf{C}^2 - 16\mathbf{C})$ and $\text{Tr}(|\log(\mathbf{C})|^2)$ are minimized by $\mathbf{C}_{1}$, however, it requires a simultaneous change of signs of all columns to transform from $\mathbf{C}_{2}$ to $\mathbf{C}_{1}$, which is not possible with a pair-wise Jacobi sweeps method. From Table~\ref{table: results}, this situation occurs only with 3/1000 probabilities. In the future, such a situation could be easily addressed with a Monte Carlo simulation.}}
    \label{table: localmin}
\end{table}
{Lastly, in table~\ref{table: localmin}, our method fails to find the global minimum of either $\text{Tr}(3\mathbf{C}^2 - 16\mathbf{C})$ or $\text{Tr}(|\log(\mathbf{C})|^2)$. Although $\text{Tr}(3\mathbf{C}^2 - 16\mathbf{C})$ is completely consistent with $\text{Tr}(|\log(\mathbf{C})|^2)$, a pair-wise Jacobi sweeps cannot transform $\mathbf{C}_{2}$ to $\mathbf{C}_{1}$, since it requires simultaneous changes of the signs of all four columns. Nevertheless, the difference is small and this situation is highly unlikely with an occurrence probability of roughly 3/1000.}

Overall, we expect that the overlap matrix should usually (i) be very close to the identity when there is no trivial crossing or (ii) have maximal element in each column to be close to $\pm1$ (in magnitude) when there is at least one trivial crossing. According to the results in table~{\ref{table: results}}, {\ref{table: signconvention}} and {\ref{table: failsignconvention}}, we conclude that our sign convention is mostly reliable and consistent, unless there are too many states crossing with each other at the same time step with strong diabatic couplings. Even in such rare cases, however, we do still recover a proper rotation matrix with a real matrix logarithm. We do not believe there should ever (in practice) be a need to reduce time step or derive higher order terms in the Taylor series on account of trivial crossings.

\nocite{*}
\bibliography{aipsamp}
\end{document}